\documentclass[12pt]{article}
\usepackage{amsfonts,amsmath,epsfig,graphicx,amssymb}
\usepackage{amsmath,epsfig,graphicx,amssymb}
\usepackage[margin=1in]{geometry}
\usepackage{amsmath}
\usepackage{amsfonts}
\usepackage{amssymb}
\usepackage{makeidx}
\usepackage{graphicx}
\newcommand{\beq}{\begin{equation}}
\newcommand{\eeq}{\end{equation}}
\newcommand{\ben}{\begin{eqnarray}}
\newcommand{\een}{\end{eqnarray}}
\date{}
\begin{document}
\title{Random Thoughts on Quantum Mechanics}
\author{Partha Ghose\footnote{partha.ghose@gmail.com} \\
Tagore Centre for Natural Sciences and Philosophy,\\ Rabindra Tirtha, New Town, Kolkata 700156, India}
\maketitle
\begin{abstract}
Here are a few random thoughts on the interpretations of the quantum double slit experiment, the Mach Zehnder experiment, the delayed-choice experiment and the measurement problem.
\end{abstract}
\section{The Quantum Double Slit Experiment}
So much has been written about the quantum double-slit experiment that one would hardly expect any further thoughts on it to be useful. Nevertheless, something struck me recently and I would like to share it with others. It's not the experiment itself or quantum mechanics as such, but interpretations of the experiment in terms of particles and waves. There's the famous exposition of it in the Feynman lectures \cite{feynman}. That's where I stumbled first. According to Feynman, if you keep only one of the slits open and send electrons (quantum mechanical particles) through it, you get a pattern similar to what one would expect from bullets ({\em Feynman Lectures}, Vol III, Chapter 1; compare the distributions $P_1$ and $P_2$, the `classical curves', in Figs 1-1 (bullets) and 1-3 (electrons)). Surely you're joking Mr. Feynman! Shouldn't you get a single-slit diffraction pattern for electrons? (See {\em Feynman Lectures}, Vol III, Chapter 2, Fig 2-2!) But in that case you see waves, don't you? If you keep both the slits open, you see a typical interference pattern ($P_{12}$ in Fig 1-3 (c)). So you see waves again! 

You may decide to send the electrons one at a time through the slits. Then you hear clicks of the detector or see single spots on the detector screen, always one at a time, {\em regardless of whether only one or both slits are open}. It's only when a sufficiently large number of electrons have gone through the apparatus that you see a pattern of spots emerge on the detector screen, either a single-slit diffraction pattern or a double slit interference pattern (with a diffraction envelope). So the pattern is always characteristic of a wave.

Finally, one talks of ``which path'' information in a double-slit set up and the consequent loss of coherence. Remember that the double-slit interference pattern is due to the overlap and superposition of the diffracted waves from the two slits, and this happens only in the `far field' region. To see which path an electron took in a given trial, one must place a detector of some kind just behind the slits and pretty close to them (Fig 1-4 in the {\em Feynman Lectures}) so that there isn't any overlap of the diffracted waves from the two slits in that region. If there is some overlap, then you can't tell which path the electron took. But if there is no overlap, you have essentially changed the apparatus from a double-slit to two single slits, and you should see two single-slit diffraction patterns! That's contextuality!  

{\flushleft{\em Question}}

So, if you take the experiment seriously, even when a single slit is open, the `quantum entity' that passes through it creates a diffraction pattern characteristic of waves but made up of discrete spots. When both slits are open, it creates an interference pattern characteristic of waves but made up of discrete spots. So, {\em there isn't any fundamental change in going from a single slit to a double slit.} Where does the Complementarity Principle come in then? 

\section{The Mach-Zehnder Interferometer and the Delayed Choice Experiment}
\begin{figure}[ht]
\centering
{\includegraphics[scale=0.7]{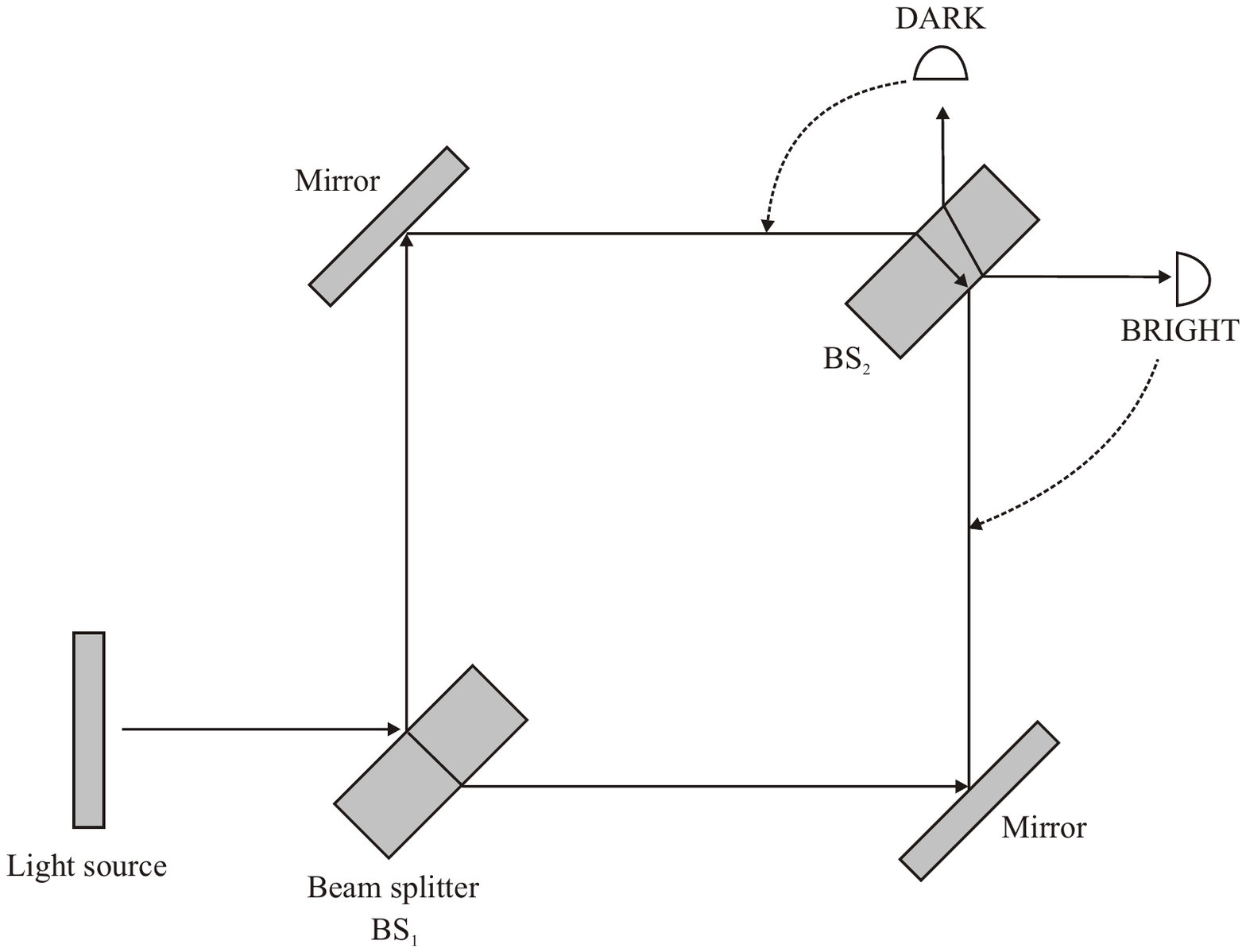}}
\caption{\label{Figure 1}{\footnotesize Mach Zehnder interferometer.}}
\end{figure}
The Mach-Zehnder interferometer offers a variant of the double-slit experiment {\em with the crucial difference that there is no diffraction in this case}. The two paths inside the interferometer after the first beam splitter $BS_1$ correspond to the paths through the two slits in a double-slit set up. If you place a detector in each path, you see a spot of light determined by the beam profile. If you don't put detectors in the paths, then after reflections by the mirrors the two beams combine on a second beam splitter $BS_2$. If the detectors are placed after $BS_2$, one of them is always dark (destructive interference) and the other is bright (constructive interference). 

If you send the particles (photons) one at a time through the interferometer, you see single detections at a time no matter where the detectors are placed (before or after $BS_2$), which don't give you any information--one swallow doesn't make a summer. You have to wait until a sufficiently large number of photons has gone through the interferometer before you can draw any conclusions. If detectors are placed in the paths after $BS_1$ but before $BS_2$, one finds spots of finite size made of individual point like spots. However, and this is important, if you watch carefully, you'll find that the detections in the two arms are always anti-coincident, telling you that the photons are indivisible quanta, or particles. Anti-coincidence on a beam splitter is a clear signal for particle-like behaviour. If they are placed after $BS_2$, one of them is always dark (no photons arrive, destructive interference) and the other bright with lots of individual point like spots (constructive interference). Here then is {\em clear evidence of both contextuality and complementarity}--particles in one case (detectors in the paths before $BS_2$) and waves in the other (detectors after $BS_2$) \cite{grangier, aspect}.

Now, it has been suggested by Wheeler that one can extend the two arms of the interferometer almost indefinitely, even right across the cosmos, and only after the `particle' has left the source and travelled a long distance from the source along one path, decide to insert the detectors just before or just after $BS_2$ (delayed-choice). If the detectors are placed before $BS_2$, one detects particles as expected, but if they are placed after $BS_2$, one detects a wave! What travelled like a particle along one path for a long while suddenly turns into a wave, depending on the delayed choice you make! So, one can indeed create what one wants to see just by changing the measurement device!

However, the quantum particle travelling along a path surely carries its phase, otherwise you can't explain the interference in the last minute. But wait, that's not the whole story. There's also the question of coherence length. You surely can't extend the paths beyond the coherence length of the wave without destroying the experiment. So, it's not clear if the experiment can be made to work at the cosmic scale.

\section{Measurements}

Let's now turn to measurement, or rather the `measurement problem' which is very interpretation specific. For example, it isn't there in Bohr's interpretation, in the many worlds interpretation or in QBism. Then where and how does it arise? Well, it's like this. If you think quantum mechanics is a universal theory applicable to everything, then a measuring apparatus is also quantum mechanical. Let a quantum mechanical system $S$ interact with a quantum mechanical measuring apparatus $A$. Then quantum mechanics tells us that the two get entangled, which means they are in a superposition of all possible quantum states $S_i$ of the system correlated with apparatus states $A_i$: $$|\Phi\rangle = \sum_i |\Phi_i\rangle = \sum_i |S_i\rangle|A_i\rangle.$$
How is it then that definite apparatus states like $|A_i\rangle$ appear in an experiment? Well, just introduce projection operators $P_i = |A_i\rangle\langle A_i|$ to project out particular states $i$ from the superposition. That converts the pure state $\rho_{SA} = |\Phi\rangle\langle \Phi|$ into a mixed state $$\hat{\rho} =\sum_i|c_i|^2\rho_i,\,\,\,\,\sum_i|c_i|^2 = 1$$ where $|c_i|^2$ are the probabilities of the various outcomes. The problem with this is the following. First of all, quantum mechanical states are intrinsically coherent, and the unitary Schr\"{o}dinger evolution preserves that coherence. But here is a process (projection or collapse) which destroys the coherence to get definite outcomes, and is therefore non-unitary. So, it can't be quantum mechanical. Furthermore, it's not a dynamical law at all! It's dictated by the outcomes! There isn't anything like that elsewhere in physics. To cap it all, the unitary Schr\"{o}dinger evolution is supposed to remain suspended during this discontinuous process of projection or collapse! These are just fiats to get the right answer. 

The problem lies in the following. One begins by making measuring apparatuses quantum mechanical against Bohr's reasoned advice not to do so, and then one has to invent some way of making the apparatuses classical back again (in the sense of preventing superpositions of states), and that's where the projection or collapse postulate comes in. Bohr made a ``{\em functional distinction} between the object and the subject of observation. This distinction is at the heart of Bohr's epistemological argument that measurement instruments lie outside the domain of the theory, insofar as they serve their purpose of acquiring empirical knowledge'' \cite{bruk}. Why then make the apparatuses quantum mechanical in the first place? Because you want to make quantum mechanics a universal theory, to encompass everything. How do you know it is? That's a postulate! No wonder no one can find a solution to the measurement problem. There isn't any.

Curiously, I happened to be in Oxford (in the summer of 1999, I think) when Freeman Dyson came to give a talk on astronomy. Roger Penrose introduced me to him before the talk, and left us in a small room for a while to finish some urgent official business. We started chatting and Dyson asked me what I was working on. When I said the foundations of quantum mechanics, he summarily brushed that aside as something quite unimportant. He said something like: `Oh, quantum mechanics was invented to account for microscopic objects. There's no reason why it should work for the macroscopic world.' I was taken aback, but fortunately Roger Penrose returned just at that point and challenged him, much to my relief. 

The crux of the problem, it seems to me, lies in the fact that we don't think of associating a wave function with a classical system. The state of a classical particle is a point in its phase space, not a vector in a Hilbert space. So you have two completely different categories of things and you can't write an interaction Hamiltonian. But there {\em is} a way of doing classical mechanics in Hilbert space, in terms of complex square integrable wave functions. That was done by Koopman \cite{K} and von Neumann \cite{vN} in the early 1930s. They introduced a wave function $\psi(q,p)$ ($[q,p] = 0$) and its complex conjugate $\psi^*(q,p)$ for a classical system, both of which obey Liouville's equation. One can then show that the phase space density defined as $\rho(q,p) = \psi^*(q,p) \psi(q,p)$ also obeys Liouville's equation, thus establishing the equivalence with classical statistical mechanics. Though the KvN wave functions are complex, their relative phases are unobservable--the phase decouples from the amplitude and there is no interference. That's inherent in the way the formalism is constructed. (For further details see \cite{ghose}). In other words, the way the formalism is set up you can't write a superposition like $\phi = \sum_i c_i \phi_i$ for KvN wave functions. The relative phases become hidden variables and one can work with just $|\phi_i|$. A superselection rule operates \cite{sud}. That takes care of the fact that classical systems don't interfere. One can still of course write a density matrix $\rho = \sum_i|c_i|^2\rho_i$, $\rho_i = \phi_i \phi_i^*$. 

Let's look at the prototype measurement of spin. Let a coherent beam of neutral spin-1/2 particles enter an inhomogeneous magnetic field (the Stern-Gerlach magnet) which turns it into the pure state 
\beq
|\psi\rangle_s = \frac{1}{\sqrt{2}}[|\uparrow\rangle_s^{(1)} + |\downarrow\rangle_s^{(2)} ] \label{a}
\eeq
where the superscripts 1,2 denote the two non-overlapping paths (ideal case). This is actually a path-spin entangled state. Only detectors placed in the two paths are relevant, and so we can write the final state as
\beq
|\Psi\rangle = \frac{1}{\sqrt{2}}[|\uparrow\rangle_s^{(1)}|\uparrow\rangle_d^{(1)} + |\downarrow\rangle_s^{(2)}|\downarrow\rangle_d^{(2)}] \label{e}
\eeq
This is a system-detector entangled state. One then applies the projection operators $P^{(1)}_\uparrow = |\uparrow\rangle_d^{(1)}\langle\uparrow|$ and $P^{(2)}_\downarrow = |\downarrow\rangle_d^{(2)}\langle\downarrow|$ on $\Psi$ to get the mixed state
\beq
\hat{\rho} = \frac{1}{2}[\rho^{(1)}_s(\uparrow)\rho^{(1)}_d(\uparrow) + \rho^{(2)}_s(\downarrow)\rho^{(2)}_d(\downarrow)].\label{b}
\eeq
If we treat the detectors as classical objects (not necessarily macroscopic), we should write KvN states for them instead of quantum mechanical states. Since the relative phases of KvN states are unobservable, a superposition like (\ref{e}) can't be written. Instead, one gets the mixed state (\ref{b}) straightaway, only it's now a quantum-classical hybrid state. No entangled state, no collapse, no measurement problem, no Wigner's friend. That's a la Bohr aided by Koopman and von Neumann.

\end{document}